\documentclass[english,pra,aps,superscriptaddress,showpacs]{revtex4}
\usepackage[T1]{fontenc}
\usepackage[latin9]{inputenc}
\usepackage{amsmath}
\usepackage{amssymb}
\usepackage{esint}
\usepackage{CJK}
\usepackage{booktabs}

\makeatletter
\@ifundefined{textcolor}{}
{%
 \definecolor{BLACK}{gray}{0}
 \definecolor{WHITE}{gray}{1}
 \definecolor{RED}{rgb}{1,0,0}
 \definecolor{GREEN}{rgb}{0,1,0}
 \definecolor{BLUE}{rgb}{0,0,1}
 \definecolor{CYAN}{cmyk}{1,0,0,0}
 \definecolor{MAGENTA}{cmyk}{0,1,0,0}
 \definecolor{YELLOW}{cmyk}{0,0,1,0}
 }

\usepackage{mathrsfs}\usepackage{amsfonts}\usepackage{epsfig}\@ifundefined{definecolor}{\@ifundefined{definecolor}
 {\usepackage{color}}{}
}{}\usepackage{CJK}\usepackage{pifont}\usepackage{bbding}

\makeatother

\usepackage{babel}

\begin{document}
\begin{CJK*}{GBK}{song}

\title{Demonstrating Additional Law of Relativistic Velocities based on
Squeezed Light}

\author{Da-Bao Yang(ÑîŽó±Š)}

\affiliation{Theoretical Physics Division, Chern Institute of Mathematics, Nankai
University, Tianjin 300071, People's Republic of China}
\affiliation{Centre for Quantum Technologies, National
University of Singapore, 3 Science Drive 2, Singapore 117543}

\author{Yan Li(ÀîÑÞ)}

\affiliation{Theoretical Physics Division, Chern Institute of Mathematics, Nankai
University, Tianjin 300071, People's Republic of China}

\author{Fu-Lin Zhang(ÕÅž£ÁÖ)}

\affiliation{Physics Department, School of Science, Tianjin University, Tianjin
300072, People's Republic of China}

\author{Jing-Ling Chen(³ÂŸ°Áé)}

\email{chenjl@nankai.edu.cn}

\selectlanguage{english}%

\affiliation{Theoretical Physics Division, Chern Institute of Mathematics, Nankai
University, Tianjin 300071, People's Republic of China}
\affiliation{Centre for Quantum Technologies, National
University of Singapore, 3 Science Drive 2, Singapore 117543}

\date{\today}
\begin{abstract}
Special relativity is foundation of many branches of modern physics,
of which theoretical results are far beyond our daily experience and
hard to realized in kinematic experiments. However, its outcomes could
be demonstrated by making use of convenient substitute, i.e. squeezed
light in present paper. Squeezed light is very important in the field
of quantum optics and the corresponding transformation can be regarded
as the coherent state of $SU(1,1)$. In this paper, the connection
between the squeezed operator and Lorentz boost is built under certain
conditions. Furthermore, the additional law of relativistic velocities
and the angle of Wigner rotation are deduced as well.

\end{abstract}


\footnotetext[1]{F.L.Z. is supported by NSF of China under Grant No. 11105097. And
J.L.C. is supported by National Basic Research Program (973 Program)
of China under Grant No. 2012CB921900 and NSF of China (Grant Nos.
10975075 and 11175089). This work is also partly supported by National
Research Foundation and Ministry of Education , Singapore (Grant No.
WBS: R-710-000-008-271).}


\section{Introduction}

\label{sec:introduction}

Special relativity is foundation of many branches of modern physics,
of which theoretical results are far beyond our daily experience and
hard to realized in kinematic experiments. Nevertheless, squeezed
light may supply a possible experimental test in modern optics laboratories.
By using canonical transformation of Wigner distribution function
in phase space, the theoretical scheme of possible experimental tests
for Wigner rotation and Thomas precession were put forward \cite{Han1988linear,Han1989thomas}.
However, the additional law of relativistic velocities, which is distinct
from the usual additional law of vectors, had not been obtained. In
present paper, we not only acquire the corresponding additional law
but also use a more direct method, which will enclose the essence
of the analogue.

As a nonclassical light field, squeezed light has a fundamental role
in the development of quantum optics, which preserves the minimum-uncertainty
product in the phase space and may exhibit many interesting properties
such as sub-poissonian photon counting statistics and photon antibunching
\cite{walls1983squeezed,fox2006quantum}. Its nonlinear generalization
and the corresponding features were studied by Kwek and Kiang \cite{kwek2003nonlinear}.
In addition, its noncyclic and nonunitary geometric phase were formulated
by Yang et. al. \cite{yang2011geometric}.

This article is organized as follows. Sec. II reviews some of indispensable
concepts of squeezed optics as well as special relativity necessary
for the present paper. In Sec. III, the connection between the squeezed
operator and Lorentz boost is built. In Sec. IV, the additional law
of relativistic velocities and the angle of Wigner rotation are deduced
as well. Moreover, a possible experimental test on the additional law
of relativistic velocities is put forward. At the end of this paper, a conclusion is drawn.

\section{reviews of squeezed state and special relativity }

\label{sec:reviews}

Squeezed state operator
\begin{equation}
S(\beta)=\exp(\frac{1}{2}\beta a^{\dagger2}-\frac{1}{2}\beta^{*}a^{2})\label{eq:SqueezedStateOperator}
\end{equation}
 can be regarded as coset space of $SU(1,1)$, which is $SU(1,1)/U(1)$,
where $\beta$ is a complex number, $*$ denotes the conjugate operation
and $a^{\dagger}$ and $a$ are creation and annihilation operators
respectively that satisfy the canonical commutation relation $[a,a^{\dagger}]=1$.
In order to demonstrate this idea clearly, let us review the necessary
knowledge about $SU(1,1)$ \cite{perelomov1986generalized}. It's
generators satisfy the following commutation relations, which are
\begin{equation}
[K_{1},K_{2}]=-iK_{0},\quad[K_{0},K_{1}]=iK_{2},\quad[K_{0},K_{2}]=-iK_{1}.\label{eq:CommutationRelations}
\end{equation}
 Via choosing another appropriate basis, the generators become
\[
K_{\pm}=\pm i(K_{1}\pm iK_{2}),\qquad K_{0}.
\]
 Therefore, the commutation relations are transformed to be
\begin{equation}
[K_{0},K_{\pm}]=\pm K_{\pm},\quad[K_{+},K_{-}]=-2K_{0}.\label{eq:WeightCommutationRelations}
\end{equation}
 Furthermore, the definition of Perelomov's $SU(1,\,1)$ coherent
state is given:
\begin{equation}
S(\beta)=\exp(\beta K_{+}-\beta^{*}K_{-}).\label{eq:CoherentState}
\end{equation}
 Last but not least, the boson realization of $SU(1,\,1)$ is supplied,
so the generators become
\begin{equation}
K_{+}=\frac{1}{2}a^{\dagger2},\quad K_{-}=\frac{1}{2}a^{2},\quad K_{0}=\frac{1}{4}(aa^{\dagger}+a^{\dagger}a).\label{eq:BosonRepresentation}
\end{equation}
which satisfy the commutation relations of \eqref{eq:WeightCommutationRelations}.
Substituting Eq. \eqref{eq:BosonRepresentation} into Eq. \eqref{eq:CoherentState},
the squeezed operator \eqref{eq:SqueezedStateOperator} is recovered.
As $SU(1,1)$ is locally isomorphic to $SO(2,1)$, the coherent state
of $SU(1,1)$ in another word the squeezed operator, may have connection
with the Lorentz boost, which will be proved in Sec. III. Moreover, the
following paragraph will review the indispensable knowledge about
special relativity.

Let us consider about $(2+1)$-dimensional spacetime. There are two
inertial frames of reference $\Sigma$ and $\Sigma^{\prime}$ that
are coincident at time $t=0$ and the velocity of $\Sigma^{\prime}$
relative to $\Sigma$ is $\textbf{v}=(v^{1},v^{2})$. An event
is observed as $(x^{0},x^{1},x^{2})^{T}$ in $\Sigma$ and $(y^{0},y^{1},y^{2})^{T}$
in $\Sigma^{\prime}$. The two teams of coordinates is connected by
a Lorentz boost $\mathcal{L}$, i.e. $(y^{0},y^{1},y^{2})^{T}=\mathcal{L}(x^{0},x^{1},x^{2})^{T}$,
where $x^{0}=ct$, $T$ denotes the operation of transposition and
$\mathcal{L}$ \cite{Ungar2001beyond} takes the form
\begin{equation}
\mathcal{L}=\left(\begin{array}{ccc}
\gamma & -\eta\gamma\frac{v^{1}}{v} & -\eta\gamma\frac{v^{2}}{v}\\
-\eta\gamma\frac{v^{1}}{v} & 1+\eta^{2}\frac{\gamma^{2}}{1+\gamma}(\frac{v^{1}}{v})^{2} & \eta^{2}\frac{\gamma^{2}}{1+\gamma}\frac{v^{1}}{v}\frac{v^{2}}{v}\\
-\eta\gamma\frac{v^{2}}{v} & \eta^{2}\frac{\gamma^{2}}{1+\gamma}\frac{v^{1}}{v}\frac{v^{2}}{v} & 1+\eta^{2}\frac{\gamma^{2}}{1+\gamma}(\frac{v^{2}}{v})^{2}
\end{array}\right),\label{eq:LorentzBoostMatrix}
\end{equation}
where $v=\sqrt{(v^{1})^{2}+(v^{2})^{2}}$, $\eta=v/c$ and $\gamma=1/\sqrt{1-\eta^{2}}.$
To simplify above Eq. \eqref{eq:LorentzBoostMatrix}, let us introduce
the auxiliary variables which are $\cosh\rho=\gamma$, $\sinh\rho=\eta\gamma$,
$\cos\varphi=v^{1}/v$ and $\sin\varphi=v^{2}/v$, so that Eq. \eqref{eq:LorentzBoostMatrix}
becomes
\begin{equation}
\left(\begin{array}{ccc}
\cosh\rho & -\sinh\rho\cos\varphi & -\sinh\rho\sin\varphi\\
-\sinh\rho\cos\varphi & 1+(\cosh\rho-1)\cos^{2}\varphi & (\cosh\rho-1)\cos\varphi\sin\varphi\\
-\sinh\rho\sin\varphi & (\cosh\rho-1)\cos\varphi\sin\varphi & 1+(\cosh\rho-1)\sin^{2}\varphi
\end{array}\right),\label{eq:SimplifiedLorentzBoostMatrix}
\end{equation}
 where the formulae $\cosh^{2}\rho-\sinh^{2}\rho=1$ and $\sinh^{2}\rho/(1+\cosh\rho)=\cosh\rho-1$
are useful. If $\varphi=0$ or $\pi/2$, readers can check that the
common Lorentz boost which always appears in the standard text book
about special relativity is reduced.

Further, let us recall additional law of relativistic velocities.
$\Sigma$, $\Sigma^{\prime}$ and $\Sigma^{\prime\prime}$ are arbitrary
three inertial frames of reference that are coincident at the initial
time. If the velocity of $\Sigma^{\prime}$ relative to $\Sigma$
is $\textbf{u}$ and the velocity of $\Sigma^{\prime\prime}$ relative
to $\Sigma^{\prime}$ is $\textbf{v}$, the velocity of $\Sigma^{\prime\prime}$
relative to $\Sigma$ is $\textbf{u}\oplus\textbf{v}$, where $\oplus$
is the additional operation of relativistic velocities which is distinct
from the parallelogram law of velocities. Its concrete formula \cite{Ungar2001beyond}
reads
\begin{equation}
\textbf{u}\oplus\textbf{v}=\frac{1}{1+\frac{\textbf{u}\cdot\textbf{v}}{c^{2}}}\left[\textbf{u}+\frac{1}{\gamma_{u}}\textbf{v}+\frac{1}{c^{2}}\frac{\gamma_{u}}{1+\gamma_{u}}(\textbf{u}\cdot\textbf{v})\textbf{u}\right].\label{eq:RelativisticAddtionLaw}
\end{equation}
 Observers at rest relative to $\Sigma$ (relative to $\Sigma^{\prime}$)
agree with observers at rest relative to $\Sigma^{\prime}$ (relative
to $\Sigma^{\prime\prime}$) that their space time coordinate are
linked by a pure Lorentz transformation without rotations. However,
if the two successive Lorentz boost are in noncollinear directions,
observers at rest relative to $\Sigma$ agree with observers at rest
relative to $\Sigma^{\prime\prime}$ that their coordinate are linked
not only by a Lorentz transformation but also by a rotation, which
is called Wigner rotation.

\section{Connecting Squeezed Transformation with Lorentz Boost}

\label{sec:Lorentz}

In this section, our main goal is to prove that the squeezed transformation
corresponds to Lorentz boost. To begin with, we express a vector
$(x^{0}\,,\, x^{1}\,,\, x^{2})$ in $(2+1)$-dimensional space time
as
\begin{equation}
x=x^{0}K_{0}-x^{1}K_{1}-x^{2}K_{2},\label{eq:Vector}
\end{equation}
 whose basis are $(K_{0},K_{1},K_{2})$, which are the generators
of $SU(1,1)$. Moreover, let us do this operation,
\begin{equation}
y^{0}K_{0}-y^{1}K_{1}-y^{2}K_{2}=S(\beta)(x^{0}K_{0}-x^{1}K_{1}-x^{2}K_{2})S^{\dagger}(\beta).\label{eq:LorentzBoostOperator}
\end{equation}
In order to get specific connection between $(y^{0},y^{1},y^{2})$
and $(x^{0},x^{1},x^{2})$, we must calculate the r.h.s. of Eq. \eqref{eq:LorentzBoostOperator}.
During the calculation, the BCH formula \cite{gerry1985dynamics}
may be useful, which takes the form
\begin{equation}
e^{A}Be^{-A}=B+[A,B]+\frac{1}{2!}[A,[A,B]]+\frac{1}{3!}[A,[A,[A,B]]]+\cdots.\label{eq:BCHFormula}
\end{equation}
 By using the above Eq. \eqref{eq:BCHFormula} and the commutations
relations \eqref{eq:CommutationRelations} of $SU(1,1)$, one can
get
\begin{equation}
\begin{array}{ccc}
S(\beta)K_{0}S^{\dagger}(\beta) & = & \cosh\rho K_{0}+\cos\varphi\sinh\rho K_{1}+\sin\varphi\sinh\rho K_{2}\\
S(\beta)K_{1}S^{\dagger}(\beta) & = & \cos\varphi\sinh\rho K_{0}+[1+(\cosh\rho-1)\cos^{2}\varphi]K_{1}\\
 &  & +(\cosh\rho-1)\cos\varphi\sin\varphi K_{2}\\
S(\beta)K_{2}S^{\dagger}(\beta) & = & \sin\varphi\sinh\rho K_{0}+(\cosh\rho-1)\cos\varphi\sin\varphi K_{1}\\
 &  & +[1+(\cosh\rho-1)\sin^{2}\varphi]K_{2}
\end{array},\label{eq:SqueezedOperationOnGenerators}
\end{equation}
 where
\begin{equation}
\beta=\frac{\rho}{2}e^{i(\pi/2-\varphi)}=\frac{\rho}{2}(\sin\varphi+i\cos\varphi).\label{eq:Condition}
\end{equation}
Substituting above Eq. \eqref{eq:SqueezedOperationOnGenerators} into
Eq. \eqref{eq:LorentzBoostOperator}, one gains that
\[
\begin{array}{cc}
 & y^{0}K_{0}-y^{1}K_{1}-y^{2}K_{2}\\
= & [\cosh\rho x^{0}-\sinh\rho\cos\varphi x^{1}-\sinh\rho\sin\varphi x^{2}]K^{0}\\
 & -\{-\sinh\rho\cos\varphi x^{0}+[1+(\cosh\rho-1)\cos^{2}\varphi]x^{1}+[(\cosh\rho-1)\cos\varphi\sin\varphi]x^{2}\}K^{1}\\
 & -\{-\sinh\rho\sin\varphi x^{0}+[(\cosh\rho-1)\cos\varphi\sin\varphi]x^{1}+[1+(\cosh\rho-1)\sin^{2}\varphi]x^{2}\}K^{2}
\end{array}.
\]
 Because the linear independence of the generators, the above equation
can be written as a matrix form, which is
\begin{equation}
\left(\begin{array}{c}
y^{0}\\
y^{1}\\
y^{2}
\end{array}\right)=\left(\begin{array}{ccc}
\cosh\rho & -\sinh\rho\cos\varphi & -\sinh\rho\sin\varphi\\
-\sinh\rho\cos\varphi & 1+(\cosh\rho-1)\cos^{2}\varphi & (\cosh\rho-1)\cos\varphi\sin\varphi\\
-\sinh\rho\sin\varphi & (\cosh\rho-1)\cos\varphi\sin\varphi & 1+(\cosh\rho-1)\sin^{2}\varphi
\end{array}\right)\left(\begin{array}{c}
x^{0}\\
x^{1}\\
x^{2}
\end{array}\right).\label{eq:SqueezedSimplifiedLorentzBoost}
\end{equation}
By comparison with Eq. \eqref{eq:SimplifiedLorentzBoostMatrix}, it
is found that the above $(3\times3)$ coefficient matrix is identical
with Lorentz boost \eqref{eq:SimplifiedLorentzBoostMatrix}. So it
has been proved that $S(\beta)xS^{\dagger}(\beta)$ is can be regarded
as Lorentz boost under the condition that $\beta=\frac{\rho}{2}(\sin\varphi+i\cos\varphi)$.
If we do the above calculations on the boson representation, the Lorentz
boost can realized by the above operation of quantum optics.

\section{Squeezed Transformation and Additional Law of Relativistic Velocities}

\label{sec:addition}

Furthermore, the next task is to demonstrate the additional law of
relativistic velocities by use of squeezed operator in this section.
We start out from this formula
\begin{equation}
S(\beta_{1})S(\beta_{2})=S(\beta_{3})R(\delta),\label{eq:WignerRotation}
\end{equation}
 where $R(\delta)=\exp(iK_{0}\delta)=\exp[i\frac{1}{4}(2a^{\dagger}a+1)\delta]$.
From the two previous sections, we get the idea that Eq. \eqref{eq:WignerRotation}
means that two Lorentz boosts is equal to not only a boost but also
a rotation. In order to represent $\beta_{3}$ and $\delta$ by use
of $\beta_{1}$ and $\beta_{2}$, let's do the following operations
on Eq. \eqref{eq:WignerRotation}, i.e. ,
\begin{equation}
S(\beta_{1})S(\beta_{2})aS(\beta_{2})^{\dagger}S(\beta_{1})^{\dagger}=S(\beta_{3})R(\delta)aR(\delta)^{\dagger}S(\beta_{3})^{\dagger}.\label{eq:BCHaa+}
\end{equation}
 With the help of the equations
\[
\begin{array}{ccc}
S(\beta)aS(\beta)^{\dagger} & = & a\cosh\frac{\rho}{2}-a^{\dagger}e^{i(\frac{\pi}{2}-\varphi)}\sinh\frac{\rho}{2}\\
S(\beta)a^{\dagger}S(\beta)^{\dagger} & = & a^{\dagger}\cosh r-ae^{-i\theta}\sinh r\\
R(\delta)aR(\delta)^{\dagger} & = & ae^{-i\frac{\delta}{2}}
\end{array},
\]
 l.h.s. and r.h.s. of Eq. \eqref{eq:BCHaa+} can be written as linear
combination of $a$ and $a^{\dagger}$ respectively. As $a$ and $a^{\dagger}$
are linear independent, the coefficient of $a$ of l.h.s. of Eq. \eqref{eq:BCHaa+}
coincides with counterpart of r.h.s., and the same is true for $a^{\dagger}$.
Hence, we obtain
\begin{equation}
a\;:\quad\cosh\frac{\rho_{1}}{2}\cosh\frac{\rho_{2}}{2}+e^{i(\varphi_{1}-\varphi_{2})}\sinh\frac{\rho_{1}}{2}\sinh\frac{\rho_{2}}{2}=e^{-i\frac{\delta}{2}}\cosh\frac{\rho_{3}}{2},\label{eq:a}
\end{equation}
 and
\begin{equation}
a^{\dagger}\;:\quad e^{-i\varphi_{1}}\sinh\frac{\rho_{1}}{2}\cosh\frac{\rho_{2}}{2}+e^{-i\varphi_{2}}\sinh\frac{\rho_{2}}{2}\cosh\frac{\rho_{1}}{2}=e^{-i\frac{\delta}{2}}e^{-i\varphi_{3}}\sinh\frac{\rho_{3}}{2}.\label{eq:aDagger}
\end{equation}
 Moreover, the above two complex Eq. \eqref{eq:a} and Eq. \eqref{eq:aDagger}
can be converted into four real equations, which are
\begin{equation}
\cosh\frac{\rho_{3}}{2}\cos\frac{\delta}{2}=\cosh\frac{\rho_{1}}{2}\cosh\frac{\rho_{2}}{2}+\cos(\varphi_{2}-\varphi_{1})\sinh\frac{\rho_{1}}{2}\sinh\frac{\rho_{2}}{2},\label{eq:aRealPart}
\end{equation}

\begin{equation}
\sin\frac{\delta}{2}\cosh\frac{\rho_{3}}{2}=\sinh(\varphi_{2}-\varphi_{1})\sinh\frac{\rho_{1}}{2}\sinh\frac{\rho_{2}}{2},\label{eq:aImaginaryPart}
\end{equation}

\begin{equation}
\cos(\theta_{3}+\frac{\delta}{2})\sinh\frac{\rho_{3}}{2}=\cos\varphi_{1}\sinh\frac{\rho_{1}}{2}\cosh\frac{\rho_{2}}{2}+\cos\varphi_{2}\sinh\frac{\rho_{2}}{2}\cosh\frac{\rho_{1}}{2}\label{eq:aDaggerRealPart}
\end{equation}
 and
\begin{equation}
\sin(\theta_{3}+\frac{\delta}{2})\sinh\frac{\rho_{3}}{2}=\sin\varphi_{1}\sinh\frac{\rho_{1}}{2}\cosh\frac{\rho_{2}}{2}+\sin\varphi_{2}\sinh\frac{\rho_{2}}{2}\cosh\frac{\rho_{1}}{2}.\label{eq:aDaggerImaginaryPart}
\end{equation}

By use of Eq. \eqref{eq:aRealPart} and Eq. \eqref{eq:aImaginaryPart},
one can reach the following results
\begin{equation}
\tan\frac{\delta}{2}=\frac{\sinh(\varphi_{2}-\varphi_{1})\sinh\frac{\rho_{1}}{2}\sinh\frac{\rho_{2}}{2}}{\cosh\frac{\rho_{1}}{2}\cosh\frac{\rho_{2}}{2}+\cos(\varphi_{2}-\varphi_{1})\sinh\frac{\rho_{1}}{2}\sinh\frac{\rho_{2}}{2}}\label{eq:WignerAngleHalf}
\end{equation}
 and
\begin{equation}
\begin{array}{ccc}
\cosh^{2}\frac{\rho_{3}}{2} & = & \sinh^{2}\frac{\rho_{1}}{2}\sinh^{2}\frac{\rho_{2}}{2}+\cosh^{2}\frac{\rho_{1}}{2}\cosh^{2}\frac{\rho_{2}}{2}\\
 &  & +2\cos(\varphi_{2}-\varphi_{1})\sinh\frac{\rho_{1}}{2}\sinh\frac{\rho_{2}}{2}\cosh\frac{\rho_{1}}{2}\cosh\frac{\rho_{2}}{2}
\end{array}.\label{eq:ConstraintLemma}
\end{equation}
 We should note that Eq. \eqref{eq:WignerAngleHalf} is just the formula
for Wigner angle \cite{mukunda2003wigner}. Substituting the following
formulae $\cosh2x=\cosh^{2}x+\sinh^{2}x$, $\cosh^{2}x-\sinh^{2}x=1$
and $\sinh2x=2\sinh x\cosh x$ into Eq.\eqref{eq:ConstraintLemma},
we can get a more elegant and pragmatic expression about $\cosh\rho_{3}$,
which is
\begin{equation}
\cosh\rho_{3}=\cosh\rho_{1}\cosh\rho_{2}+\cos(\varphi_{2}-\varphi_{1})\sinh\rho_{1}\sinh\rho_{2}.\label{eq:Constraint}
\end{equation}
 In addition, let us introduce the following auxiliary variables
\begin{equation}
\begin{array}{ccccccccccc}
\gamma_{u} & = & \cosh\rho_{1} &  & \gamma_{v} & = & \cosh\rho_{2} &  & \gamma_{w} & = & \cosh\rho_{3}\\
\textbf{u} & = & c\tanh\rho_{1}\hat{u} &  & \textbf{v} & = & c\tanh\rho_{2}\hat{v} &  & \textbf{w} & = & c\tanh\rho_{3}\hat{w}
\end{array},\label{eq:Gamma}
\end{equation}
 where $\hat{u}$, $\hat{v}$ and $\hat{w}$ are unit vectors along
the direction of $\textbf{u}$, $\textbf{v}$ and $\textbf{w}$ respectively
and $c$ is the velocity of light propagating in the vacuum. Substituting
Eq. \eqref{eq:Gamma} into Eq. \eqref{eq:Constraint}, we can get
a more meaningful formula, which is
\begin{equation}
\gamma_{w}=\gamma_{u}\gamma_{v}(1+\frac{\textbf{u}\cdot\textbf{v}}{c^{2}})\label{eq:AdditionOfGamma}
\end{equation}

Now, we concentrate on another two equations \eqref{eq:aDaggerRealPart}
and \eqref{eq:aDaggerImaginaryPart}, by substituting $\cos\delta/2$
from Eq. \eqref{eq:aRealPart} and $\sin\delta/2$ from Eq. \eqref{eq:aImaginaryPart}
into them, then they compose a set of equations about unknown elements
$\cos\varphi_{3}$ and $\sin\varphi_{3}$, i.e. ,
\begin{equation}
\left(\begin{array}{cc}
A & B\\
C & D
\end{array}\right)\left(\begin{array}{c}
\cos\varphi_{3}\\
\sin\varphi_{3}
\end{array}\right)=\left(\begin{array}{c}
E\\
F
\end{array}\right),\label{eq:LinearEquations}
\end{equation}
 where $A=-D=\tanh\frac{\rho_{3}}{2}\sin(\varphi_{2}-\varphi_{1})\sinh\frac{\rho_{1}}{2}\sinh\frac{\rho_{2}}{2}$,
$B=C=\tanh\frac{\rho_{3}}{2}[\cosh\frac{\rho_{1}}{2}\cosh\frac{\rho_{2}}{2}+\cos(\varphi_{2}-\varphi_{1})\sinh\frac{\rho_{1}}{2}\sinh\frac{\rho_{2}}{2}]$,
$E=\cos\varphi_{1}\sinh\frac{\rho_{1}}{2}\cosh\frac{\rho_{2}}{2}+\cos\varphi_{2}\sinh\frac{\rho_{2}}{2}\cosh\frac{\rho_{1}}{2}$
and $F=\sin\varphi_{1}\sinh\frac{\rho_{1}}{2}\cosh\frac{\rho_{2}}{2}+\sin\varphi_{2}\sinh\frac{\rho_{2}}{2}\cosh\frac{\rho_{1}}{2}$.
According to Cramer's rules, the above Eq. \eqref{eq:LinearEquations}
can be solved, hence we can obtain
\begin{equation}
\cos\varphi_{3}=\frac{\left|\begin{array}{cc}
E & B\\
F & D
\end{array}\right|}{\left|\begin{array}{cc}
A & B\\
C & D
\end{array}\right|}=\begin{array}{c}
\frac{1}{\sinh\rho_{3}}\{[\sinh\rho_{1}\cosh\rho_{2}+\sinh\rho_{2}(\cosh\rho_{1}-1)\times\\
(\cos\varphi_{1}\cos\varphi_{2}+\sin\varphi_{1}\sin\varphi_{2})]\cos\varphi_{1}+\sinh\rho_{2}\cos\varphi_{2}\}
\end{array}\label{eq:xDirection}
\end{equation}
 and
\begin{equation}
\sin\varphi_{3}=\frac{\left|\begin{array}{cc}
A & B\\
C & D
\end{array}\right|}{\left|\begin{array}{cc}
A & B\\
C & D
\end{array}\right|}=\begin{array}{c}
\frac{1}{\sinh\rho_{3}}\{[\sinh\rho_{1}\cosh\rho_{2}+\sinh\rho_{2}(\cosh\rho_{1}-1)\times\\
(\cos\varphi_{1}\cos\varphi_{2}+\sin\varphi_{1}\sin\varphi_{2})]\sin\varphi_{1}+\sinh\rho_{2}\sin\varphi_{2}\}
\end{array}.\label{eq:yDirection}
\end{equation}
 In order to disclose the meaning of additional law of relativistic
velocities. Let us introduce three unit vectors, which are
\[
\hat{u}=\cos\varphi_{1}\hat{i}+\sin\varphi_{1}\hat{j}\quad\hat{v}=\cos\varphi_{2}\hat{i}+\sin\varphi_{2}\hat{j}\; and\hat{\; w}=\cos\varphi_{3}\hat{i}+\sin\varphi_{3}\hat{j}.
\]
 Hence Eq. \eqref{eq:xDirection} and Eq. \eqref{eq:yDirection} become
\begin{equation}
\hat{w}=\frac{1}{\sinh\rho_{3}}\{[\sinh\rho_{1}\cosh\rho_{2}+\sinh\rho_{2}(\cosh\rho_{1}-1)\hat{u}\cdot\hat{v}]\hat{u}+\sinh\rho_{2}\hat{v}\}.\label{eq:AdditionOfUnitVectors}
\end{equation}
 Moreover, substituting Eq. \eqref{eq:Gamma} and Eq. \eqref{eq:AdditionOfGamma}
into Eq. \eqref{eq:AdditionOfUnitVectors}, the final addition law
\begin{equation}
\textbf{w}=\frac{1}{1+\frac{\textbf{u}\cdot\textbf{v}}{c^{2}}}(\textbf{u}+\frac{\gamma_{u}}{\gamma_{u}+1}\frac{\textbf{u}\cdot\textbf{v}}{c^{2}}\textbf{u}+\frac{1}{\gamma_{u}}\textbf{v})\label{eq:RelativisticAddtionLawFinal}
\end{equation}
 is achieved, which is identical with Eq. \eqref{eq:RelativisticAddtionLaw}.
When $c\rightarrow\infty$, the common additional law of velocities
\[
\textbf{w}=\textbf{u}+\textbf{v}
\]
 in Galilean transformation is obtained.

Depending on the analysis above, the additional law of relativistic velocities
has been successfully achieved from squeezed optics. Furthermore, the theoretical
results can be possibly verified by the following experiment. In practice, it is much easier to
realize the two-mode squeezed state
\[
S^{\rm tm}(\beta)=\exp[\frac{1}{2}(\beta a_{1}^{\dagger}a_{2}^{\dagger}+\beta^{*}a_{1}a_{2})]
\]
than one-mode one, where subscripts $1$ and $2$ denote the two different
photons and superscript ``tm" represents two-mode, thus we
choose two-mode process. Because the generators of two-mode squeezed
operator obey the same Lie algebra as the one-mode squeezed operator,
our analysis above also applies. At first, a beam of light was splitted
into two beams. Second, beam $1$ goes through a nondegenerate optical
parametric amplifier (NOPA) which could generate two-mode squeezed
vacuum state \cite{reid1998quantum}. As a result, beam $1$ becomes
$S^{\rm tm}(\beta_{2})|00\rangle$. Moreover, it continues to pass an equipment
which makes the light become $S^{\rm tm}(\beta_{1})S^{tm}(\beta_{\rm 2})|00\rangle$.
Up to beam $2$, it is changed to be $S^{\rm tm}(\beta_{3})|00\rangle$
by NOPA, where $\beta_{3}$ is calculated from $\beta_{1}$ and $\beta_{2}$
according to Eq. \eqref{eq:Constraint}, \eqref{eq:xDirection} and
\eqref{eq:yDirection}. After all the transformations on beams $1$
and $2$, let them interfere with each other. Then the dark and bright
fringes can be observed. The separation of adjacent bright fringes
gives $\delta^{\prime}$. If $\delta^{\prime}$ coincides with $\delta$
of Eq. \eqref{eq:WignerAngleHalf}, then the additional law of relativistic velocities is demonstrated.

\section{Conclusion and acknowledgments}

\label{sec:conclusion}

In summary, since squeezed transformations constitute coset space
of $SU(1,1)$, Lorentz transformations make up Lorentz group and $SU(1,1)$
group is locally isomorphic to the $(2+1)$-dimensional Lorentz group
\cite{perelomov1986generalized}, we demonstrate the phenomenon of
special relativity by use of squeezed optics. At first it is proved
that the squeezed transformation \eqref{eq:LorentzBoostOperator}
is equivalent to Lorentz boost \eqref{eq:SqueezedSimplifiedLorentzBoost}
under the condition \eqref{eq:Condition}. Furthermore, the additional
law of relativistic velocities \eqref{eq:RelativisticAddtionLawFinal}
and the angle of Wigner rotation \eqref{eq:WignerAngleHalf} are deduced
as well. Specifically speaking, the relations between squeezed parameters
($\beta_{1}=\frac{\rho_{1}}{2}e^{i(\pi/2-\varphi_{1})}$, $\beta_{2}=\frac{\rho_{2}}{2}e^{i(\pi/2-\varphi_{2})}$
and $\beta_{3}=\frac{\rho_{3}}{2}e^{i(\pi/2-\varphi_{3})}$ in Eq.
\eqref{eq:WignerRotation}) and velocities ($\textbf{u}$, $\textbf{v}$
and $\textbf{w}$ in Eq. \eqref{eq:RelativisticAddtionLawFinal}) are
illustrated below,
\[
\begin{array}{c|cc}
\toprule[1pt]
velocity & modulo & directional\: vector\\
\midrule[0.5pt] \textbf{u} & c\tanh\rho_{1} & (\cos\varphi_{1},\sin\varphi_{1})\\
\textbf{v} & c\tanh\rho_{2} & (\cos\varphi_{2},\sin\varphi_{2})\\
\textbf{w} & c\tanh\rho_{3} & (\cos\varphi_{3},\sin\varphi_{3})\\
\bottomrule[1pt]
\end{array}.
\]
Moreover, a possible experimental test on the additional law of relativistic
velocities is also discussed.

\bibliographystyle{plain}

\end{CJK*}
\end{document}